\documentstyle[preprint,aps]{revtex}
\begin{document}
\draft
\preprint{KWUTP-94/1}

\title{Type II Anyon Superconductivity}

\author{Jewan Kim and Jungdae Kim}
\address{Department of Physics, Seoul National University, Seoul,151-742,
KOREA}
\author{Taejin Lee}
\address{Department of Physics, Kangwon National University, Chuncheon 200-701,
KOREA }

\maketitle

\begin{abstract}

Assuming that the superconductivity which is described by the low energy
effective action of the anyon system may be type II, we discuss its
characteristics. We also study physical
properties of the Chern-Simons vortices, which may be formed as the external
magnetic field is applied to the system, such as their statistics, the free
energy of single vortex and the interaction energy between two vortices.

\end{abstract}

\pacs{PACS numbers: 74.20.-z,74.65.+n,11.15.Kc}

\narrowtext

In a seminal paper \cite{fhl} Fetter, Hanna and Laughlin made an important step
in the anyon superconductivity \cite{lsw}, showing that a neutral
gas of anyons has a massless Goldstone mode and exhibits the Meissner effect.
Their work has been further extended by Chen, Halperin, Wilczek and Witten
\cite{chww}. Banks and Lykken \cite{bl} then subsequently found that it is
possible to describe the system in terms of a standard Landau-Ginzburg order
parameter if the radiatively induced Chern-Simons term cancels its bare
counterpart. However, there are many subjects to be clarified before the anyon
superconductivity is accepted as a realistic model for the high-$T_c$
superconductivity. One of them is the issue raised by the recent observation
\cite{high} that all known high-$T_c$ superconductors are of type II: The issue
is whether the anyon superconductivity can be extended to describe the type II
superconductivity. The purpose of this paper is to discuss the anyon
superconductivity in the framework of the type II superconducitvity theory and
to study physical properties of the Chern-Simons vortices which play an
important role in the type II anyon superconductivity.

In the Abelian-Higgs model with the Chern-Simons term \cite{ahcs} there
exist vortex solutions of finite energy as in the case where the Chern-Simons
term is absent. However, these Chern-Simons vortices have
distinctive characters: Carrying both electric charge and magnetic flux, they
acquire fractional spins and satisfy exotic statistics. The Chern-Simons term
\cite{djt} introduces the Aharonov-Bohm \cite{ab} interaction between the
vortices,  endowing them with the magnetic fluxes. In this respect, the effect
of the Chern-Simons term in the 2+1 dimensional Abelian-Higgs model is similar
to that of the CP violating term $\theta {F}^*F$ in the four dimensional
non-Abelian-Higgs model which leads to the dyon \cite{dyon}. The Chern-Simons
vortex persists \cite{hkp,jw} even in the Chern-Simons-Higgs model \cite{csh}
where the Maxwell term is suppressed. Moreover, in such a case the vortex
solutions can be self-dual, satisfying the Bogomol'nyi conditions \cite{bogo},
provided that the Higgs potential is of a certain form.

The anyon superconductor theory may be described by the Lagrangian density
\cite{bl}
\begin{equation}
L=-{1\over 4}F_{\mu\nu}F^{\mu\nu}-\Pi
f_{\mu\nu}f^{\mu\nu}+\kappa\epsilon^{\mu\nu\lambda}A_\mu \left(F_{\nu\lambda}
+2 f_{\nu\lambda}\right) \label{lag}
\end{equation}
where $\Pi$ and $\kappa$ are constants and $a_\mu$ is the statistical gauge
field, $f_{\mu\nu}=\partial_\mu a_\nu-\partial_\nu a_\mu$.
Its non-relativistic version has been discussed in refs.\cite{fhl,chww}.
The Lagrangian density may be derived as a low-energy effective action for
the system of anyons with the electromagnetic interaction
\begin{equation}
L = -{1\over 4}F_{\mu\nu}F^{\mu\nu} +\kappa^\prime
\epsilon^{\mu\nu\lambda}a_\mu f_{\nu\lambda} +\bar\psi\left(D\cdot
\gamma+m\right) \psi \label{lagany}
\end{equation}
where $D_\mu = \partial_\mu +ie(A_\mu-a_\mu)$.
The relativistic effective Lagrangian for the fermion field $\psi$ may
result from some microscopic model such as that of ref.\cite{haldane}.
Integrating out the fermion field $\psi$ results in the effective action
\begin{equation}
-i \ln {\rm Det} \left(D\cdot \gamma+m\right),
\end{equation}
which has the low-energy expansion \cite{djt,oneloop} given by
\begin{equation}
-\frac{1}{48\pi} \frac{e^2}{|m|}\left(F_{\mu\nu}-f_{\mu\nu}\right)
\left(F^{\mu\nu}-f^{\mu\nu}\right) +\frac{e^2}{16\pi}\frac{m}{|m|}
\epsilon^{\mu\nu\lambda} \left(A_\mu-a_\mu\right)
\left(F_{\nu\lambda}-f_{\nu\lambda}\right) +\dots . \label{exp}
\end{equation}
We may remove $F^{\mu\nu}F_{\mu\nu}$ and $f^{\mu\nu} F_{\mu\nu}$ terms
in the above low-energy expansion by shifting the statistical gauge field,
$a_\mu \rightarrow a_\mu+A_\mu$. Thus, the low-energy effective action
for the system of anyons with the electromagnetic interaction is written
as $\int L_{\rm eff}$ where
\begin{equation}
L_{\rm eff} = -{1\over 4}F_{\mu\nu}F^{\mu\nu} -\frac{1}{48\pi}
\frac{e^2}{|m|} f_{\mu\nu}f^{\mu\nu} + \frac{e^2}{16\pi}\frac{m}{|m|}
\epsilon^{\mu\nu\lambda}a_\mu f_{\nu\lambda} +\kappa^\prime
\epsilon^{\mu\nu\lambda}\left(A_\mu+a_\mu\right)
\left(F_{\nu\lambda}+f_{\nu\lambda}\right).\label{lageff}
\end{equation}
{}From the effective action Eq.(\ref{lageff}) we may read the condition
\cite{bl}
for the massless Goldstone mode to exist: The radiatively induced Chern-Simons
term for the statistical gauge field cancels its bare term
\begin{equation} \kappa^\prime
= -\frac{e^2}{16\pi}\frac{m}{|m|}. \label{cond}
\end{equation}
This cancellation, which would not occur if $\psi$ is nonrelativistic,
holds up exactly all higher order due to the non-renormalization
theorem \cite{nren1,nren2} for the Chern-Simons term.
Note that the fermion $\psi$ has a statistical phase $\theta =\pm 2\pi$
with the above value of $\kappa^\prime$, so the Chern-Simons interaction
does not alter the statistics of $\psi$. Comparison of Eq.(\ref{lag}) and
Eq.(\ref{lageff}) yields the two constants $\Pi$ and $\kappa$
\begin{equation}
\Pi = \frac{1}{48\pi} \frac{e^2}{|m|},
\qquad \kappa = -\frac{e^2}{16\pi}\frac{m}{|m|}.\label{const}
\end{equation}

The field equations of motion which result from the Lagrangian density
Eq.(\ref{lag}) are
\begin{mathletters}
\label{eqm:all}
\begin{eqnarray}
\partial_\nu F^{\mu\nu}-2\kappa
\epsilon^{\mu\nu\lambda}\left(F_{\nu\lambda} +f_{\nu\lambda}\right) & = & 0,
\label{eqm:a}\\
\Pi\partial_\nu f^{\mu\nu}-\kappa\epsilon^{\mu\nu\lambda}\partial_\nu
A_\lambda & = & 0. \label{eqm:b}
\end{eqnarray}
\end{mathletters}
Eq.(\ref{eqm:b}) is satisfied if we take
\begin{equation}
f^{\mu\nu}={\kappa\over \Pi}\epsilon^{\mu\nu\lambda}A_\lambda+c
\epsilon^{\mu\nu\lambda}\partial_\lambda\phi
\end{equation}
where $c$ is a non-zero constant and will be chosen ${1\over 2\Pi}$ for
convenience. Then the equations of motion are rewritten as
\begin{mathletters}
\label{eqmr:all}
\begin{eqnarray}
\partial_\nu F^{\mu\nu}-2\kappa
\epsilon^{\mu\nu\lambda}F_{\nu\lambda} -{2\kappa\over
\Pi}\left(\partial^\mu\phi+2\kappa A^\mu\right) & =&0, \label{eqmr:a}\\
\partial_\mu\left(\partial^\mu\phi+2\kappa A^\mu\right) & =&0.\label{eqmr:b}
\end{eqnarray}
\end{mathletters}
Eq.(\ref{eqmr:b}) is the Bianchi identity
\begin{equation}
{\partial_\mu}^*f^\mu=\partial_\mu\left({1\over 2}\epsilon^{\mu\nu\lambda}
f_{\nu\lambda}\right)=0.
\end{equation}

One easily identifies Eq.(\ref{eqmr:all}) as the equations of motion which
follow from the Lagrangian
\begin{equation}
L=-{1\over 4}F_{\mu\nu}F^{\mu\nu}+\kappa\epsilon^{\mu\nu\lambda}
A_\mu F_{\nu\lambda}+{1\over 2\Pi}\left(\partial^\mu\phi+2\kappa
A^\mu\right)^2.\label{lagsup}
\end{equation}
The obtained Lagrangian Eq.(\ref{lagsup}) is similar to that of the well-known
Landau-Ginzburg model \cite{lg} or equivalently the Abelian-Higgs model. If it
were not for the Chern-Simons term, it would be
considered as the Abelian-Higgs model
\begin{equation}
L=-{1\over 4}F_{\mu\nu}F^{\mu\nu}+{1\over 2}\left\vert\left(\partial_\mu+
i2eA_\mu\right)\chi\right\vert^2+V(\vert\chi\vert)\label{lagah}
\end{equation}
in the London limit \cite{london} where the expectation value of
$\vert\chi\vert$ is kept constant. The London limit is particularly useful to
describe the vortex of type II superconductor \cite{abri,type2} in that
the coherence length $\xi$ characterizing the size of the core of the vortex is
much smaller than other characteristic lengths, and the variation
of the order parameter $\vert\chi\vert$ is negligible outside of the core.
Note that, however, the vortex solutions apparently look singular in the London
limit. This observation suggests that, although the vortex solutions to be
explored in what follows are singular, they may be relevant to
the anyon superconductivity.
Therefore, it is meaningful to study the vortex solution of the Lagrangian
Eq.(\ref{lagsup}) and to discuss their roles in the anyon superconductivity in
the context of the type II superconductor theory. Recalling that the most
physical properties of the vortex in the type II superconductor can be well
described without invoking the inside structure of the core, we expect that
we get a fairly good description of the type II anyon superconductor.

For a static
solution the equations of motion Eq.(\ref{eqmr:all}) reduce to
\begin{mathletters}
\label{sta:all}
\begin{eqnarray}
\partial_j F^{ij}-4\kappa\epsilon^{ij}\partial_j A_0-{2\kappa\over \Pi}
\left(\partial^i\phi+2\kappa A^i\right)& =&0  \label{sta:a} \\
\nabla^2 A_0-{4\kappa^2\over \Pi}A_0+4\kappa H & =&0 \label{sta:b} \\
\partial_i\left(\partial^i\phi+2\kappa A^i\right) &=&0 \label{sta:c}
\end{eqnarray}
\end{mathletters}
where $H=-{1\over 2}\epsilon^{ij} F_{ij}$.
Choosing the Coulomb gauge $\partial^i A_i=0$, we set $A_i=\epsilon_{ij}
\partial^j f$. Taking the Ansatz for the  axially symmetric vortex
solution of vorticity $n$
\begin{equation}
A_0=A_0(r),\quad f=f(r),\quad {\rm and}\,\,\,\phi={\kappa\over e
}n\theta, \label{ansa}
\end{equation}
we can further reduce the equations of motion to
\begin{mathletters}
\label{red:all}
\begin{eqnarray}
\left(\nabla^2-\mu^2\right)f-4\kappa A_0 & =&\left({n\mu^2\over
2e}\right){\rm ln}r \label{red:a} \\
\left(\nabla^2-\mu^2\right)A_0-4\kappa\nabla^2 f & =&0  \label{red:b}
\end{eqnarray}
\end{mathletters}
where $\mu^2={4\kappa^2\over \Pi}$.
The reason to take the third condition in the Ansatz Eq.(\ref{ansa}) is that we
may identify $\phi$ as a Goldstone mode, i.e.,
\begin{equation}
\chi=\vert\chi\vert e^{i{e\over \kappa}\phi},
\end{equation}
which follows from the comparison of Eq.(\ref{lagsup}) with Eq.(\ref{lagah}).
Concurrently we find that
\begin{equation}
|\chi| = \frac{|\kappa|}{e\sqrt{\Pi}}= \sqrt{\frac{3|m|}{16\pi}}.
\end{equation}

With some algebra we can solve exactly the differential equations
(\ref{red:all})
\begin{mathletters}
\label{exac:all}
\begin{eqnarray}
f(r)&=&-\frac{n}{4e(\Pi+1)}\left[ \frac{1}{1+\Delta}
\left(K_0(a_+r)+\ln r\right)+ \frac{1}{1-\Delta}
\left(K_0(a_-r)+\ln r\right)\right] \label{exac:a} \\
A_0(r)&=& -\frac{\kappa n \Delta}{2 e \Pi}
\left[ K_0(a_+r)-K_0(a_-r)\right] \label{exac:b}
\end{eqnarray}
\end{mathletters}
where $a_\pm= 2|\kappa| \left(\Delta^{-1} \pm 1\right)$ and
$\Delta = \sqrt{\frac{\Pi}{\Pi+1}}<1$. The
electric charge density associated with the vortex is simply $j_0=\mu^2A_0$.
Here $K_0$ is the modified Bessel function which we represent as
\begin{equation}
 K_0\left(\nu\vert{\bf x}-{\bf x}^\prime\vert\right)=-\int{d^2k\over 2\pi}
{e^{-i{\bf k}\cdot({\bf x}-{\bf x}^\prime)}\over k^2+\nu^2}.
\end{equation}
The total magnetic flux $\Phi$ and the electric charge $Q$ of the vortex which
the solution Eq.(\ref{exac:all}) describes are
\begin{mathletters}
\label{vor:all}
\begin{eqnarray}
\Phi &=&\int d^2x H = -2\pi r{\partial f\over \partial
r}\biggr|_{r=\infty}={\pi n\over e},  \label{vor:a} \\
Q & =&\int d^2x \mu^2 A_0=-\frac{ne}{4}\frac{m}{|m|}
=4\kappa \Phi. \label{vor:b}
\end{eqnarray}
\end{mathletters}
The relation between $\Phi$ and $Q$ can be read from the equations of motion,
so it applies to general static solutions. According to Eq.(\ref{vor:all})
and Eq.(\ref{const}) the
vortex itself is an anyon with the statistical parameter
\begin{equation}
\theta_s=\frac{Q\Phi}{2}= -\frac{n^2}{8} \frac{m}{|m|}\pi.\label{vorst}
\end{equation}
Note that the statistical parameter for the vortex of vorticity $n$
depends upon the charge of $\chi$ which we assign to it. If we choose the
charge
of $\chi$ to be $pe$, the statistical parameter for the vortex would read as
\[ \theta_s = -\frac{n^2}{2p^2} \frac{m}{|m|}\pi. \]
We choose here $2e$ for the charge of $\chi$, because we consider $\chi$
corresponds to the bound state of a pair of $\psi$.

Making use of the analytic exact expression Eq.(\ref{exac:all}), we can
evaluate
the electric and magnetic fields that the vortex carries
\begin{mathletters}
\label{em:all}
\begin{eqnarray}
{\bf E}({\bf x}) & =&{{\bf x}\over r} \frac{\kappa^2 n}{2\Pi}
\left[ (1+\Delta) K_1(a_+r) - (1-\Delta) K_1 (a_-r)\right],  \label{em:a} \\
H({\bf x}) & =& -\nabla^2 f(r)  \nonumber \\
 & =& \frac{\kappa^2 n}{e\Pi}\left[ (1+\Delta)
K_0(a_+r)+(1-\Delta)K_0(a_-r)\right].\label{em:b}
\end{eqnarray}
\end{mathletters}
By means of the asymptotic expressions of the modified Bessel function
$$K_0(x)=-{\rm ln}\,{x\over 2}-\gamma\quad {\rm for}\quad x\ll 1,\qquad
K_0(x)=\sqrt{\pi\over 2x}e^{-x}\quad {\rm for} \quad x\gg 1,$$
we can read the behaviors of the various physical quantities near the core or
far from it. In particular, we find that the magnetic field behaves as
\begin{equation}
H({\bf x})
= \frac{\kappa^2 n}{e\Pi}\sqrt{\frac{\pi}{2r}}\left[
\frac{(1+\Delta)}{\sqrt{a_+}} e^{-a_+r} +
\frac{(1-\Delta)}{\sqrt{a_-}} e^{-a_-r}\right]
\end{equation}
for $r\gg \delta_\pm > \xi$, so the decrease of the field is
characterized by two penetration depth parameters $\delta_\pm$.
The analytic expression Eq.(\ref{exac:all}) also enables us to evaluate the
free
energy (or the mass of the vortex) of a single vortex with vorticity $n$
\begin{eqnarray}
F_n &=&\int d^2x\left({1\over 2}(H^2+E^2)+{1\over 2\Pi}\left(4\kappa^2
A^2_0+\left(\partial^i\phi+2\kappa A^i\right)^2\right)\right)  \nonumber \\
  &=& \frac{1}{2\pi}\left(\frac{e n\Delta}{16\Pi}\right)^2
\left[\frac{1}{1+\Delta} \ln\left(\delta_+/\xi\right)+\frac{1}{1-\Delta}
\ln\left(\delta_-/\xi\right) \right].\label{mass} \end{eqnarray}
Here we assume that the contribution from the inside
region of the core is negligible. (This is true if the inside of the core is
described by Ginzburg-Landau type Lagrangian.)
Note that the coherence length $\xi$ cannot be determined by the Lagrangian
Eq.(\ref{lagsup}) and may be fixed by some higher powers of fields which are
suppressed in the Lagrangian Eq.(\ref{lag}). Here we take it as a
phemenological
parameter characterzing the theory. However, since  $F_n$ has logarithmic
accuracy, it is insensitive to the precise value of $\xi$.
Equation (\ref{mass}) indicates that the mass $F_n$ is proportional to $n^2$ as
in the case of the usual type II superconductor vortex, therefore, formation
of vortex lines with one flux quantum is the most favourable.

The interaction energy of two
vortices at a distance $d\gg \xi$ can be calculated similarly. Thanks to the
linearity of the equations of motion Eq.(\ref{sta:all}) which are valid outside
of the core, the system of two vortices at a distance $d\gg \xi$ can be
described
by a superposition of the fields of individual vortices. To evaluate the
interaction energy $F_{12}$ of two vortices we calculate the total free energy
of the system, then substract the masses of the individual vortices from it.
Using the equations of motion Eq.(\ref{sta:all}), we rewrite the free energy as
\begin{eqnarray}
F(H,A_0)&=&{1\over 2}\int d^2 x\left(H^2+E^2+\mu^2 A^2_0+{1\over
\mu^2}\left(\partial_i\left(H+4\kappa A_0\right)\right)^2\right)  \nonumber \\
  &=&{1\over 2}\int_S d{\bf l}\cdot {{\bf \nabla} \over 2}\left(A^2_0+{1\over
\mu^2}\left(H+4\kappa A_0\right)^2\right)
\end{eqnarray}
where $S$ is the boundary of the two dimensional space. For the system under
discussion the total free energy is given by $F(H^T,A^T_0)$ with $H^T({\bf
r})=H({\bf r}-{\bf r}_1)+H({\bf r}-{\bf r}_2)$, $A^T_0({\bf r})=A_0({\bf
r}-{\bf r}_1)+A_0({\bf r}-{\bf r}_2)$ and $S$ composed of two small circles,
say, $S_1$ and $S_2$, enclosing the cores of the vortices. Here
$A_0({\bf r})$ and $H({\bf r})$ are the fields describing a
single vortex located at ${\bf r}$ of which explicit expressions are given by
Eq.(\ref{exac:b}) and Eq.(\ref{em:b}). Accordingly the interaction energy
$F_{12}$ is  given by
\begin{eqnarray}
F_{12}(d) &=&{1\over 2}\int_{S_1} d{\bf l}\cdot {\bf
\nabla}\left(A_0+{1\over \mu^2}\left(H+4\kappa A_0\right)\right)({\bf r}-{\bf
r}_1)   \nonumber \\
  &   & \qquad
\left(A_0+{1\over \mu^2}\left(H+4\kappa A_0\right)\right)({\bf r}-{\bf
r}_2)+(1\leftrightarrow 2)  \nonumber  \\
 & =& \frac{1}{\pi} \left(\frac{e n}{16}\right)^2
\frac{(1-\Delta^2)}{\Delta^2} \left[(1-\Delta)
K_0(a_+d) + (1+\Delta) K_0(a_-d)\right].
\end{eqnarray}
The interaction is always repulsive as in the ususal case of the type II
superconductor.

A few brief comments are in order on the critical fields of the anyon
superconductivity model. The clear distinction
between type I and type II superconductors lies in their magnetic properties.
A typical type II superconductor has two critical fields, $H_{c1}$
and $H_{c2}$: When the applied magnetic field $H$ is smaller than $H_{c1}$,
the superconductor exhibits the perfect Meissner effect. As the applied
field exceeds $H_{c1}$, the vortices begin to form and for $H_{c1} < H
< H_{c2}$ the specimen enters a new state called mixed state where a lattice of
supercurrent vortices is formed. Finally for $H > H_{c2}$ the specimen
becomes normal. The critical field $H_{c1}$ can be determined by making use of
the analytic expression of the free energy of a single vortex Eq.(\ref{mass}).
If the free energy of the single $F_n$ vortex is smaller than
the magnetic energy $F_M$, the formation of a vortex is favourable.
The vortex of the vorticity $n$ first appears when
\begin{equation}
F_n = \Phi_n H_{c1}.
\end{equation}
This yields $H_{c1}$
\begin{equation}
H_{c1} = \frac{eF_n}{\pi n}
\simeq \frac{3e n|m|}{32\pi} \left(\ln (\delta_+/\xi)
+\ln (\delta_-/\xi)\right).
\end{equation}
But the critical field $H_{c2}$ cannot be determined by the given
Lagrangian Eq.(\ref{lag}). In order to evaluate it we may need the terms of
higher powers in fields which are suppressed in the effective action.

We conclude this paper with some remarks on what remains to be discussed
further.
In this paper we explored the effective action for the anyon superconductor in
the context of the type II superconductivity theory and showed that it is
certainly possible for the anyon superconductor to be type II. However, there
are several subjects to be clarified in the future works.
We expect that the magnetization curve of the anyon superconductor near the
critical field $H_{c1}$ may radically differ from that of regular type II
superconductor. The reason is that there are two penetration depth parameters
instead of one and the vortex of the lowest free energy satisfies the
fractional statistics rather than the bose statistics.
Our discussion of this paper shows that it is urgent to investigate
the question of whether the terms of higher powers in fields may
improve the theory (e.g. they render
the vortex solutions regular) and help us to determine $\xi$ and $H_{c2}$.
Last but not least, it is also important to clearly understand the effect
of the four-fermi interaction \cite{4fermi}, in the anyon
superconductor theory, which plays an essential role in the ordinary
BCS type superconductor theory.

\acknowledgments
This work was supported in part by non-directed research fund,
Korea Research Foundation (1992) and in part by BSRI-92-206.

\end{document}